\documentclass[showpacs,aps,prx,floatfix,reprint,superscriptaddress,footinbib]{revtex4-2}
\usepackage{amssymb,amsfonts,amsmath}
\usepackage{graphicx}
\usepackage{verbatim}
\usepackage[utf8]{inputenc}
\usepackage[T1]{fontenc}
\usepackage{soul,xcolor,colortbl}
\usepackage{multirow}
\usepackage{bm}
\usepackage{array}
\usepackage{color}
\usepackage{hyperref} \hypersetup{colorlinks=true,citecolor=blue,linkcolor=blue,urlcolor=blue}
\usepackage{mdframed}
\usepackage{lineno}

\definecolor{pranab_green}{rgb}{0.31,0.53,0.10}
\definecolor{pranab_red}{rgb}{0.85,0.23,0.11}

\newcolumntype{L}[1]{>{\raggedright\arraybackslash}p{#1} }
\newcolumntype{C}[1]{>{\centering  \arraybackslash}p{#1} }
\newcolumntype{R}[1]{>{\raggedleft \arraybackslash}p{#1} }

\def\AFLOW{{\small AFLOW}}

\def\AFLOWPOCC{{\small AFLOW-POCC}}

\def\EFA{{\small EFA}}
\def\SQS{{\small SQS}}

\makeatletter \renewcommand\frontmatter@abstractwidth{\dimexpr\textwidth\relax} \makeatother

\begin{document}
\title{High-entropy ceramics: propelling applications through disorder}

\author{Cormac Toher}
\affiliation{Dept. Mechanical Engineering and Materials Science and Center for Autonomous Materials Design, Duke University, Durham, NC 27708, USA}
\author{Corey Oses}
\affiliation{Dept. Mechanical Engineering and Materials Science and Center for Autonomous Materials Design, Duke University, Durham, NC 27708, USA}
\author{Marco Esters}
\affiliation{Dept. Mechanical Engineering and Materials Science and Center for Autonomous Materials Design, Duke University, Durham, NC 27708, USA}
\author{David Hicks}
\affiliation{Dept. Mechanical Engineering and Materials Science and Center for Autonomous Materials Design, Duke University, Durham, NC 27708, USA}
\author{George N. Kotsonis}
\affiliation{Dept. Materials Science and Engineering, The Pennsylvania State University, University Park, PA 16802, USA}
\author{Christina M. Rost}
\affiliation{Dept. Physics and Astronomy, James Madison University, Harrisonburg, VA 22807, USA}
\author{Donald W. Brenner}
\affiliation{Dept. Materials Science and Engineering, North Carolina State University, Raleigh, NC 27695, USA}
\author{Jon-Paul Maria}
\affiliation{Dept. Materials Science and Engineering, The Pennsylvania State University, University Park, PA 16802, USA}
\author{Stefano Curtarolo}
\email[]{stefano@duke.edu}
\affiliation{Dept. Mechanical Engineering and Materials Science and Center for Autonomous Materials Design, Duke University, Durham, NC 27708, USA}

\date{\today}

\begin{abstract}
\noindent
Disorder enhances desired properties, as well as creating new avenues for synthesizing materials.
For instance, hardness and yield stress are improved by solid-solution strengthening, a result of distortions and atomic size mismatches.
Thermo-chemical stability is increased by the preference of chemically disordered mixtures for high-symmetry super-lattices.
Vibrational thermal conductivity is decreased by force-constant disorder without sacrificing mechanical strength and stiffness.
Thus, high-entropy ceramics propel a wide range of applications: from wear resistant coatings and thermal and environmental barriers to catalysts, batteries, thermoelectrics and nuclear energy management.
Here, we discuss recent progress of the field, with a particular emphasis on disorder-enhanced properties and applications.
\end{abstract}
\maketitle

\section*{Introduction}

\noindent
Entropy is expected to dominate the formation of single-phase multicomponent ceramics~\cite{curtarolo:art152},
opening up new synthesis routes and facilitating property optimization without the need for toxic or expensive elements.
High-entropy ceramics~\cite{Oses_NatureReview_2020} typically consist of a disordered multi-cation sublattice and an ordered single-anion sublattice~\cite{curtarolo:art99}, with multi-anion ceramics now also being developed~\cite{Qin_BoroCarbide_JEurCeramSoc_2020,Dippo_CarboNitride_SciRep_2020, Wen_CarboNitride_JAmCeramSoc_2020}.
The earliest work on high-entropy ceramics involved thin films synthesized by sputtering high-entropy alloys in a N$_2$ or O$_2$ atmosphere~\cite{Chen_Nitride_SurfCoatTech_2004, Chen_Nitride_SurfCoatTech_2005, Lai_Nitride_SurfCoatTech_2006}.
The original attempts typically resulted in amorphous or multi-phase films due to the presence of non-nitride forming metals such as Cu~\cite{Chen_Nitride_SurfCoatTech_2004, Chen_Nitride_SurfCoatTech_2005};
while the first single-phase high-entropy ceramic thin film reported was (AlCrTaTiZr)N, displaying an fcc lattice based on the rocksalt structure~\cite{Lai_Nitride_SurfCoatTech_2006}.
In 2015, (MgCoNiCuZn)O became the first bulk single-phase high-entropy ceramic to be synthesized --- it also displayed a rocksalt-type fcc lattice~\cite{curtarolo:art99}, and was demonstrated to be entropy-stabilized:
when annealed at high temperatures, the non-cubic CuO and ZnO secondary phases merged with the rocksalt phase~\cite{curtarolo:art99}.
High-entropy ceramics have now expanded to include carbides~\cite{curtarolo:art140}, borides~\cite{Gild_borides_SciRep_2016}, silicides~\cite{Gild_Silicides_JoM_2019}, niobates~\cite{Zhu_HENiobates_JEurCeramSoc_2021, Chen_HENiobates_ApplPhysLett_2021}, zirconates~\cite{Zhou_HEZirconate_JEurCeramSoc_2020, Zhu_HEZirconate_JEurCeramSoc_2021}, as well as multi-anion boro-carbides~\cite{Qin_BoroCarbide_JEurCeramSoc_2020} and carbo-nitrides~\cite{Dippo_CarboNitride_SciRep_2020, Wen_CarboNitride_JAmCeramSoc_2020}.

In addition to contributing to synthesizability, disorder directly alters and improves the properties of materials such as high-entropy ceramics~\cite{Oses_NatureReview_2020}.
Solid-solution strengthening combined with reduced grain growth enhances mechanical properties including strength, hardness and toughness~\cite{curtarolo:art140, curtarolo:art148, Csanadi_HECarbide_JEurCeramSoc_2020, Moskovskikh_Nitride_SciRep_2020} ---  vital for structural applications and wear resistant coatings~\cite{Braic_CarbideNitride_JMechBehavBiomedMater_2012}.
The strong thermodynamic preference for high-symmetry lattices increases the thermo-chemical stability, suppressing structural defect formation and phase transitions --- crucial for thin-film diffusion barriers in nanoelectronics~\cite{Tsai_Nitride_ApplPhysLett_2008} and for improving cycling ability in battery materials~\cite{Sarkar_LiEnergyHEO_NComm_2018, Chen_HEBatteries_ACSApplMaterInt_2021} ---
while entropic stabilization of unusual oxidation states opens up catalyst design rules~\cite{Fracchia_HEOCatalysis_JPhysChemLett_2020}.
Interatomic force constant and mass disorder reduce lattice thermal conductivity, often reaching the amorphous limit~\cite{Braun_ESO_AdvMat_2018} --- useful for increasing the efficiency of thermoelectric devices~\cite{Zheng_HEPerovskite_JAdvCeram_2021, Jiang_HEChalcogenides_Science_2021, Jian_HEChalcogenides_NatCommun_2021}.
The combination of low thermal conductivity, mechanical strength and corrosion resistance make high-entropy ceramics particularly suitable as thermal and environmental barriers (\textit{e.g.}, for hypersonics) ~\cite{Zhu_HENiobates_JEurCeramSoc_2021, Chen_HENiobates_ApplPhysLett_2021,Zhou_HEZirconate_JEurCeramSoc_2020, Zhu_HEZirconate_JEurCeramSoc_2021}.
In magnetic systems, disorder introduces localization, reducing the orbital overlap and increasing the importance of double and superexchange~\cite{Sharma_HEMagnetic_PhysRevMater_2020},
while disrupting long-range order promotes the formation of relaxor ferroelectrics~\cite{Pu_Ferroelectric_ApplPhysLett_2019, Liu_HEferroelectric_CeramInt_2020, Zhang_HEFerroelectric_MaterDes_2021}.

\section*{Synthesis and modeling techniques}

\noindent
\textbf{Synthesis approaches.}
Synthesis techniques for high-entropy ceramics include sputtering~\cite{Chen_Nitride_SurfCoatTech_2004, Chen_Nitride_SurfCoatTech_2005, Lai_Nitride_SurfCoatTech_2006}; pulsed laser deposition~\cite{Kotsonis_HEO_PhysRevMater_2020}; (reactive) spark plasma sintering~\cite{curtarolo:art140, Qin_HEBorides_JEurCeramSoc_2020};  boro-, carbo- and boro-carbothermal synthesis~\cite{Feng_JACerS_2020}; sol-gel synthesis~\cite{Fracchia_HEOCatalysis_JPhysChemLett_2020}; and solution combustion synthesis~\cite{Saghir_HEOSCS_JEurCeramSoc_2021}.
One current research direction is developing techniques to avoid grain-coarsening to enhance mechanical properties, such as new spark-plasma sintering protocols with lower temperatures or shorter annealing times \cite{Zhang_HECarbides_JMaterSciTech_2021},
or carbothermal synthesis using finely ground metal or metal oxide precursors~\cite{Feng_JACerS_2020, Feng_HECarbide_JAmCerSoc_2019}.
Growth kinetics (\textit{e.g.}, substrate temperature for pulsed laser deposition) and epitaxial strain can be used to engineer electrical and magnetic properties~\cite{Kotsonis_HEO_PhysRevMater_2020}.
Rapid high-temperature Joule heating was used to synthesize a Pd-containing denary nanoparticle catalyst for methane combustion~\cite{Li_NCat_2021},
where short, high-temperature heat treatments overcame the kinetic barriers for mixing multiple elements while avoiding agglomeration and collapse of pore structure.

\noindent
\textbf{Computational modeling.}
First-principles modeling of disordered materials often uses special quasi-random structures (\SQS)~\cite{zunger_sqs}, in which the lattice is decorated so as to maximize the similarity of the radial correlation functions to that of the completely disordered material, in combination with the ideal approximation for the configurational entropy.
\SQS\ models the fully disordered, infinite-temperature limit: useful for understanding how high levels of disorder can affect properties, but neglecting the effect of short- and long-range ordering at finite temperatures.

The \AFLOW\ \underline{p}artial \underline{occ}upation module (\AFLOWPOCC) enables the modeling of disorder as a function of temperature~\cite{curtarolo:art110}.
The disordered material is represented by an ensemble of small ordered cells, generated by enumerating all possible decorations of the parent structure that produce the target compositions up to a cell-size cut-off.
A descriptor, the \underline{e}ntropy \underline{f}orming \underline{a}bility (\EFA), was formulated based on the energy distribution obtained from first-principles calculations of these configurations~\cite{curtarolo:art140}.
A narrow distribution implies that it is easy to introduce new configurations and disorder,
while a broad distribution indicates a thermodynamic preference for certain ordered states.
The \EFA\ descriptor has guided the synthesis of high-entropy carbides, successfully predicting whether compositions would form a single-phase or undergo phase separation~\cite{curtarolo:art140}.
 Cr-containing compositions forming a single phase were also identified with machine-learning models trained on \EFA\ data~\cite{curtarolo:art164}.

The properties of high-entropy ceramics have also been modeled computationally~\cite{curtarolo:art140, Sangiovanni_ElasticHEC_MaterDes_2021, Dai_ThermoElasticBorides_JMaterSciTech_2021}.
\AFLOWPOCC\ was used to calculate the thermodynamically weighted average of the elastic moduli of the different configurations to predict the mechanical properties of disordered carbides~\cite{curtarolo:art140}.
The temperature-dependent elastic properties of (HfTaTiWZr)C and (MoNbTaVW)C and their component binary carbides were investigated using \textit{ab initio} molecular dynamics~\cite{Sangiovanni_ElasticHEC_MaterDes_2021}.
Molecular dynamics simulations with deep-learning potentials were used to predict the thermal and elastic properties of (HfNbTaTiZr)B$_2$ as a function of temperature~\cite{Dai_ThermoElasticBorides_JMaterSciTech_2021}.

\section*{Disorder-enhanced properties}

\begin{figure}
\includegraphics[width=0.99\linewidth]{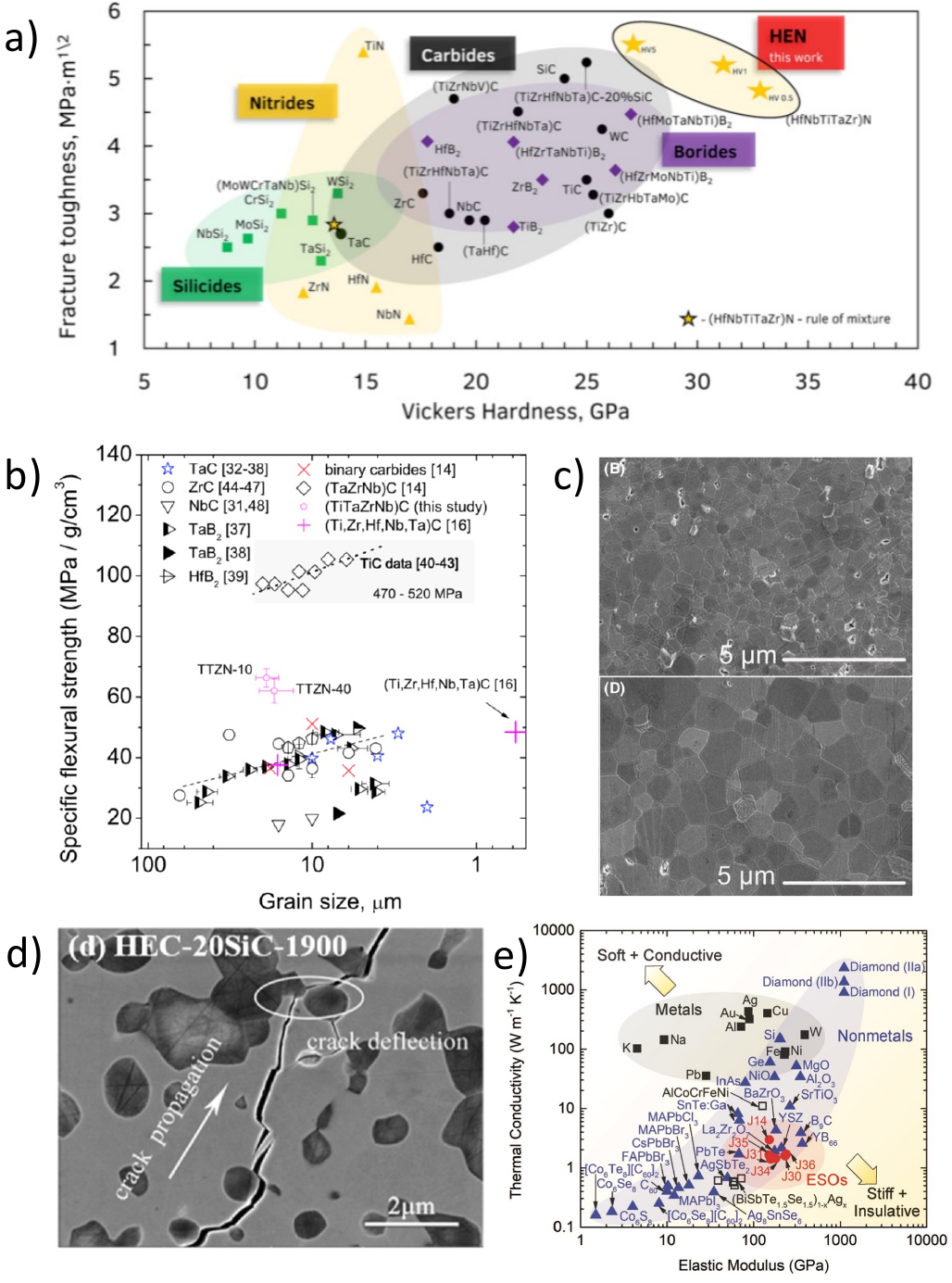}
\vspace{-3mm}
\caption{\small \textbf{Thermal and mechanical properties.}
\textbf{a)} Disorder simultaneously enhances hardness and fracture toughness of high-entropy nitrides (panel adapted from Ref. \onlinecite{Moskovskikh_Nitride_SciRep_2020}).
\textbf{b)} Flexural strength as a function of grain size (panel adapted from Ref. \onlinecite{Demirskyi_HECarbides_OpenCeram_2020}).
\textbf{c)} Microstructures of (HfNbTaTiZr)C annealed at 1800$^\circ$C (top) and 1900$^\circ$C (bottom), showing the increase in grain size with temperature (panel adapted from Ref. \onlinecite{Feng_HECarbide_JAmCerSoc_2019}).
\textbf{d)} Crack deflection by SiC phase (panel adapted from Ref. \onlinecite{Lu_HECarbidesSiC_JEurCeramSoc_2020}).
\textbf{e)} Thermal conductivity \textit{versus} elastic modulus: disorder lowers thermal conductivity while maintaining stiffness (panel adapted from Ref. \onlinecite{Braun_ESO_AdvMat_2018}).}
\label{fig:thermal_mechanical}
\end{figure}

\noindent
\textbf{Mechanical properties.}
The mechanical properties of high-entropy ceramics are enhanced by a combination of grain refinement and solid-solution strengthening ---
lattice distortions due to differing atomic radii and Jahn-Teller effects~\cite{Rost_HEOx_JACerS_2017, Rak_HEO_JT2_MatLett_2018} create barriers that disrupt the propagation of dislocations, hindering plastic deformation and increasing hardness and yield strength.
The reduction in grain coarsening has been attributed to increased crystalline energy due to lattice distortions, which reduces the free energy gained by shrinking the grain surface area~\cite{Huang_Nitride_ScrMater_2010, Wang_GrainSizeHEC_JAmCeramSoc_2020}, as well as to slow diffusion~\cite{Wang_GrainSizeHEC_JAmCeramSoc_2020}.
Finer grains lead to improved toughness, and to Hall-Petch strengthening due to the nanograin microstructure impeding the motion of dislocations.

Bulk (HfNbTaTiZr)C, prepared by spark-plasma sintering, shows a hardness 50\% higher than what would be expected from the rule-of-mixtures of the binary carbides~\cite{curtarolo:art140, curtarolo:art148, Ye_Carbide_JAmCerS_2018},
while carbo-nitride versions of this composition are slightly harder than the carbides~\cite{Dippo_CarboNitride_SciRep_2020, Wen_CarboNitride_JAmCeramSoc_2020}, and an 8-cation carbide is even harder~\cite{Wang_HECarbides_AdvTheorySimul_2020}.
High-entropy nitrides show even greater improvements, doubling both hardness and fracture toughness compared to their components (Figure~\ref{fig:thermal_mechanical}a)~\cite{Moskovskikh_Nitride_SciRep_2020}.
Thin films deposited using magnetron sputtering had even higher hardness~\cite{Mayrhofer_Boride_ScrMater_2018,Gorban_Carbide_JSuperhardMater_2017, Liang2_Nitride_ApplSurfSci_2011}, likely due to internal stresses induced by ion bombardment~\cite{Mayrhofer_Boride_ScrMater_2018, Gorban_Carbide_JSuperhardMater_2017, Liang2_Nitride_ApplSurfSci_2011}: the highest reported hardness was 70~GPa for a (HfNbTaTiV)N film~\cite{Sobol_Nitride_TechPhysLett_2012}.
The highest fracture toughness values of 8.4~MPa m$^{1/2}$ were reported for (MoNbTaTiVW)C(N) \cite{Peng_Carbides_ApplPhysLett_2019},
whereas the highest flexural strengths tend to be for compositions with fewer elements such as (NbTaTiZr)C (544 to 560~MPa~\cite{Demirskyi_HECarbides_OpenCeram_2020}) or (TaTiZr)C at 700~MPa~\cite{Demirskyi_MechCarbide_JAsCeramSoc_2020}.

The enhanced mechanical properties of high-entropy ceramics are generally maintained to relatively high temperatures~\cite{Demirskyi_MechCarbide_JAsCeramSoc_2020, Feng_JACerS_2020, Han_CreepHEC_JEurCeramSoc_2020}.
(TaTiZr)C retained half its flexural strength up to 1800$^\circ$C, while its fracture toughness increased almost 40\% when heated to 1800$^\circ$C~\cite{Demirskyi_MechCarbide_JAsCeramSoc_2020}.
The flexural strength of a fine-grained (HfNbTaTiZr)C sample remained almost constant up to ${\sim}1800^\circ$C, before decreasing significantly between 2000$^\circ$C and 2300$^\circ$C \cite{Feng_JACerS_2020}.
The steady state creep rates of (HfNbTaZr)C are about 10 times lower than the rates for the component binary carbides~\cite{Han_CreepHEC_JEurCeramSoc_2020}, attributed to lattice distortion and sluggish diffusion.

Synthesis procedures, defects and grain size all have a major impact on mechanical properties of high-entropy materials  (Figure~\ref{fig:thermal_mechanical}b for dependence of strength on grain size)~\cite{Csanadi_HECarbide_JEurCeramSoc_2020, Hossain_VacanciesHEC_ActaMater_2021, Zhang_HECarbides_JMaterSciTech_2021, Wang_GrainSizeHEC_JAmCeramSoc_2020}.
The fracture strength of (HfNbTaZr)C was significantly reduced by defects in the form of pores or secondary phases~\cite{Csanadi_HECarbide_JEurCeramSoc_2020}.
Reducing the number of carbon vacancies in high-entropy carbides increases the hardness up to carbon saturation~\cite{Hossain_VacanciesHEC_ActaMater_2021}, while excess carbon precipitates at the grain boundaries, first reducing hardness and then increasing it again due to the formation of a diamond-like carbon matrix.
Average grain size in (HfMoNbTaTiZr)C increased ${\sim}4$-fold with increase in sintering temperature up to 2500$^\circ$C~\cite{Zhang_HECarbides_JMaterSciTech_2021},
reducing the Vickers hardness.
Sub-micron grains of (HfNbTaTiZr)C were prepared using a two-step sintering process~\cite{Wang_GrainSizeHEC_JAmCeramSoc_2020}:
the small grain size was thermally stable, and the sample showed improved strength, hardness and toughness.
Fine-grained high-entropy carbide samples can also be prepared by carbothermal synthesis of oxide precursors, with grain size controlled by the hot-pressing temperature (Figure~\ref{fig:thermal_mechanical}c)~\cite{Feng_JACerS_2020}.

Mechanical properties of high-entropy ceramics can be further enhanced by incorporating SiC secondary phases: SiC particles deflect cracks and increase toughness (see Figure~\ref{fig:thermal_mechanical}d)~\cite{Lu_HECarbidesSiC_JEurCeramSoc_2020, Liu_HEBoridesSiC_JAdvCeram_2020}.
In particular, 4-point bending strengths increased by 40\% to 60\%, while fracture toughness improved by about 20\%.

\noindent
\textbf{Thermal conductivity.}
Disorder suppresses lattice thermal conductivity in high-entropy ceramics without compromising mechanical strength and stiffness (Figure~\ref{fig:thermal_mechanical}e)~\cite{Braun_ESO_AdvMat_2018, Rost_HECarbide_ActaMater_2020, Liu_HECarbide_JEurCeramSoc_2020, Wright_PyrochloreThermalConductivity_ScrMater_2020}.
(MgCoNiCuZn)O has a lattice thermal conductivity of $2.95\pm0.25$~W~m$^{-1}$~K$^{-1}$, while six-cation systems formed by adding other elements such as Sc, Sb, Sn, Cr or Ge have even lower values.
The lowest thermal conductivities for high-entropy carbides were for porous samples~\cite{Chen_Carbide_JMaterSciTech_2019} --- when fully densified, thermal conductivities range from ${\sim}2$~W~m$^{-1}$~K$^{-1}$ to 9.2~W~m$^{-1}$~K$^{-1}$~\cite{Yan_2018_5metalC_thermal_conductivity, Liu_HECarbide_JEurCeramSoc_2020, Rost_HECarbide_ActaMater_2020}.
Reducing the carbon content of films increased the metallic bonding, raising the electronic contribution to the thermal conductivity \cite{Rost_HECarbide_ActaMater_2020}.
Very low thermal conductivities were reported for multi-component chalcogenides~\cite{Jiang_HEChalcogenides_Science_2021, Jian_HEChalcogenides_NatCommun_2021} --- promising for thermoelectric applications.

The reduction in thermal conductivity has been attributed to disorder in the interatomic force constants~\cite{Braun_ESO_AdvMat_2018} and atomic radii~\cite{Wright_PyrochloreThermalConductivity_ScrMater_2020},
while mass disorder appears to be less significant~\cite{Braun_ESO_AdvMat_2018}.
Extended x-ray absorption fine structure measurements show that the oxygen sublattice distortion is much larger in 6-cation systems such as (MgNiCuCoZnCr)O than in (MgCoNiCuZn)O~\cite{Braun_ESO_AdvMat_2018}, indicating the importance of lattice distortion and force constant disorder --- this is further supported by molecular dynamics simulations~\cite{Braun_ESO_AdvMat_2018}.
For a set of 22 oxides based on the pyrochlore structure, the thermal conductivity had a good correlation ($r^2 = 0.73$) with the overall size disorder~\cite{Wright_PyrochloreThermalConductivity_ScrMater_2020}.

\begin{figure}
\includegraphics[width=0.99\linewidth]{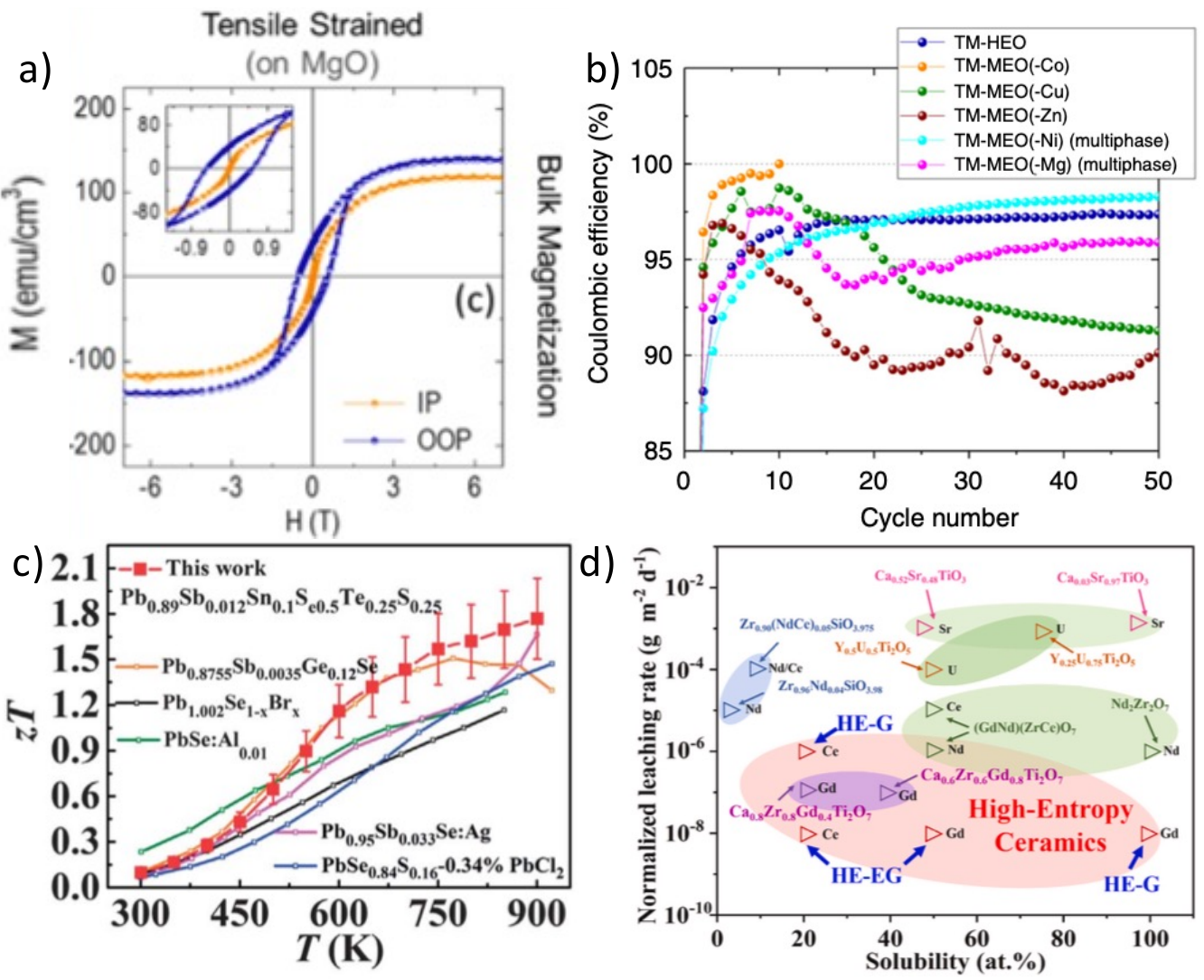}
\vspace{-3mm}
\caption{\small \textbf{High-entropy ceramics for magnets, batteries, thermoelectrics and radioactive waste immobilization.}
\textbf{a)} High-entropy oxide thin films showed strong magnetic anisotropy when subjected to in-plane epitaxial strain (panel adapted from Ref. \onlinecite{Sharma_HEMagnetInsulator_ACSApplMaterInt_2021}).
\textbf{b)} High-entropy battery materials maintain efficiency under cycling (panel adapted from Ref. \onlinecite{Sarkar_LiEnergyHEO_NComm_2018}).
\textbf{c)} Increasing the number of components improves the thermoelectric efficiency  (panel adapted from Ref. \onlinecite{Jiang_HEChalcogenides_Science_2021}).
\textbf{d)} High-entropy ceramics show reduced leaching rates when used as immobilization matrices for radioactive waste (panel adapted from Ref. \onlinecite{Zhou_RadioactiveWastePyrochlore_JHazMater_2021}).}
\label{fig:magnet_battery_thermoelectric_radiation}
\end{figure}

\noindent
\textbf{Electrical and magnetic properties.}
The low long-range order in high-entropy ceramics strongly affects their electrical and magnetic properties.
Ferroelectric and magnetic domains tend to be smaller and the number of microstates is increased.
Frustrated dipolar and magnetic ordering can promote the formation of ``relaxor ferroelectrics''~\cite{Pu_Ferroelectric_ApplPhysLett_2019, Liu_HEferroelectric_CeramInt_2020, Zhang_HEFerroelectric_MaterDes_2021} and spin glasses~\cite{Meisenheimer_ESO_PhysRevMater_2019}.
In magnetic systems, disorder introduces localization, reducing orbital overlap and raising the importance of double and superexchange~\cite{Sharma_HEMagnetic_PhysRevMater_2020}.
Entropy also enables the fabrication of thermodynamically stable single-phase magnetic insulators with high quality surfaces such as (MgNiFeCoCu)Fe$_2$O$_4$~\cite{Sharma_HEMagnetInsulator_ACSApplMaterInt_2021}: promising for spintronic, spincaloritronic, nonvolatile memory and microwave applications.

The magnetic and electrical properties of high-entropy ceramics can be tuned by altering their composition~\cite{Jacobson_HEOConductivity_JEurCeramSoc_2021, Meisenheimer_ESO_PhysRevMater_2019} or synthesis conditions~\cite{Jacobson_HEOConductivity_JEurCeramSoc_2021, Kotsonis_HEO_PhysRevMater_2020}, or by applying strain~\cite{Sharma_HEMagnetic_PhysRevMater_2020, Sharma_HEMagnetInsulator_ACSApplMaterInt_2021}.
(MgCoNiCuZn)O films grown at low temperature exhibit enhanced Co$^{3+}$ concentrations that result in a compressed lattice parameter, reduced optical band gap, increased electrical conductivity, and more diluted magnetic lattice ~\cite{Kotsonis_HEO_PhysRevMater_2020, Jacobson_HEOConductivity_JEurCeramSoc_2021}.
Tensile strain generates magnetic anisotropy by increasing the out-of-plane magnetic hardness (Figure~\ref{fig:magnet_battery_thermoelectric_radiation}a)~\cite{Sharma_HEMagnetic_PhysRevMater_2020}, apparently due to the modification of the superexchange coupling between the transition metal atoms.
Increasing the Cu content of (MgCoNiCuZn)O creates structural disorder since Cu$^{2+}$ ions tend to undergo a tetragonal distortion from the octahedral coordination (Jahn-Teller effect~\cite{Rak_HEO_JT2_MatLett_2018}),
leading to increased spin frustration and a reduction in the driving force for a preferred magnetization axis~\cite{Meisenheimer_ESO_PhysRevMater_2019}.
Increasing the non-magnetic Co$^{3+}$ concentration in (MgCoNiCuZn)O results in more uncompensated spins when interfaced with a ferromagnet, leading to enhanced magnetic exchange bias~\cite{Kotsonis_HEO_PhysRevMater_2020}.

Relaxor ferroelectrics have much smaller permanent electric dipole domains than conventional ferroelectrics, forming ``polar nano-regions'' on the nanoscale instead of the microscale.
They take less energy to align, giving them very high specific capacitance and making them useful for energy storage.
They also have high dielectric permittivity, high piezoelectric coefficient and giant field induced strain: important for applications in dielectric capacitors, piezoelectric sensors and actuators.
In high-entropy ceramics such as perovskite-structure (NaBiBaSrCa)TiO$_3$ and  (BiNaKBaCa)TiO$_3$, or Aurivillius phase (CaSrBaPb)Bi$_2$Nb$_2$O$_9$ and (Ca$_{0.2}$Sr$_{0.2}$Ba$_{0.2}$Pb$_{0.2}$Nd$_{0.1}$Na$_{0.1}$)Bi$_2$Nb$_2$O$_9$ (space group  $A2_1am$, \#36), differences in the valence and ion size disrupt long-range ferroelectric ordering and result in the formation of several different types of electric dipoles.
This promotes the formation of polar nano-regions, reducing the saturation polarization, remanent polarization and coercive field \cite{Pu_Ferroelectric_ApplPhysLett_2019, Liu_HEferroelectric_CeramInt_2020, Zhang_HEFerroelectric_MaterDes_2021}, and increasing the piezoelectric coefficients \cite{Zhang_HEFerroelectric_MaterDes_2021}.
(NaBiBaSrCa)TiO$_3$ also displayed an electrocaloric effect: the application of an electric field changes the dipole arrangement from disordered to ordered, resulting in an entropy-change that can be exploited for refrigeration \cite{Pu_Ferroelectric_ApplPhysLett_2019}.

\noindent
\textbf{Corrosion resistance.}
The high melting temperatures of carbides and borides offer great potential for applications in extreme environments such as combustion chambers, where resistance against corrosion is critical.
Oxidation behavior has been extensively studied in (HfNbTaTiZr)C~\cite{Ye_Chu_CorSci_2019,Ye_Chu_JACERS_2020,Tan_Zhang_JAC_2020,Backmann_Opila_ActaMat_2020_Part_I,Backmann_Opila_ActaMat_2020_Part_II}.
Kinetic investigations showed that oxidation follows a parabolic rate law between 1073 and 1773~K, indicating a diffusion-controlled process~\cite{Ye_Chu_CorSci_2019,Ye_Chu_JACERS_2020,Tan_Zhang_JAC_2020}.
The rate constant increases until 1273~K while the passivating layer is highly porous, then decreases with increasing temperature as more complex quaternary and quinary oxides form and the passivating layer densifies ~\cite{Ye_Chu_CorSci_2019,Ye_Chu_JACERS_2020}.
The oxide layer is heterogeneous: group IV oxides primarily form near the surface (due to the readily oxidized corresponding carbides), while group V oxides form near the interface with the ceramic~\cite{Backmann_Opila_ActaMat_2020_Part_I,Backmann_Opila_ActaMat_2020_Part_II} --- similar results were found in the diboride analog~\cite{Backmann_Opila_ActaMat_2020_Part_I,Backmann_Opila_ActaMat_2020_Part_II}.
Experiments using water vapor showed a denser protective layer, which increased corrosion resistance by preventing inward diffusion of H$_2$O~\cite{Tan_Zhang_JAC_2020}.
The same study found improved corrosion resistance compared to ZrC --- attributed to sluggish diffusion in the high-entropy material due to lattice distortions, suppressing outward diffusion of the cations.
This was supported by decreased oxidation rates in (HfNbTaTiZr)C compared to (TiNbTaZr)C and (NbTiZr)C.

A separate study in air found improved oxidation resistance in (HfNbTaZr)C compared to the five-metal carbide~\cite{Wang_Reece_CorSci_2020,Wang_Wang_JMST_2021}.
The four-metal carbide also showed improved performance compared to the binary carbide reactants, and the oxide layer was more structurally stable than in (HfTa)C, even though (HfTa)C had a lower oxidation onset temperature~\cite{Wang_Reece_CorSci_2020}.
Mechanistic studies confirmed that outward diffusion of TiO is the rate-determining step in the oxidation of (HfNbTaTiZr)C and thus key to increased corrosion resistance of (HfNbTaZr)C~\cite{Wang_Wang_CERI_2020}.
Performance was further improved by doping with SiC as it forms a dense oxide that prevents inwards diffusion of oxygen.
Additionally, SiO is an effective diffusion barrier for Ta and Nb oxides, preventing outward diffusion of these metals~\cite{Wang_Wang_CERI_2020,Wang_Wang_JMST_2021}.
(HfNbTaZr)C with 20 vol-\% SiC showed enhanced oxidation resistance compared to the ultra-high temperature ceramic ZrB$_2$-SiC~\cite{Wang_Wang_JMST_2021}.
Composition and porosity of the passivating oxide layer are therefore crucial to the stability of high-entropy ceramics at high temperatures under oxidizing conditions.
The ability to combine many elements to tailor chemical properties makes high-entropy carbides and borides strong candidates for applications in extreme heat conditions.

\section*{High-entropy ceramic applications}

\noindent
\textbf{Batteries.}
The entropy-driven preference for high-symmetry lattices gives materials such as (MgCoNiCuZn)O high thermo-chemical stability, suppressing structural deformation and defect formation, giving them the potential to maintain performance under charge cycling (Figure~\ref{fig:magnet_battery_thermoelectric_radiation}b) \cite{Sarkar_LiEnergyHEO_NComm_2018, Qiu_HEOLi_JAC_2019}.
This led to investigations of the suitability of high-entropy ceramics for use as anodes, cathodes and solid-state electrolytes in Li-ion batteries \cite{Sarkar_LiEnergyHEO_NComm_2018, Chen_HEBatteries_ACSApplMaterInt_2021}.
Shortly after its initial synthesis, room temperature superionic conductivity was reported for Li-doped (MgCoNiCuZn)O \cite{Berardan_2016_JMCA_ESO_superionic}, significantly exceeding that of the currently-used lithium phosphorous oxy-nitride solid-state electrolyte.

High-entropy oxides based on (MgCoNiCuZn)O are also promising as anodes, with high specific capacities and robust responses to charge cycling.
Initial capacities of 980~mAh~g$^{-1}$ \cite{Sarkar_LiEnergyHEO_NComm_2018} to 1585~mAh~g$^{-1}$ \cite{Qiu_HEOLi_JAC_2019} have been reported, with a reversible capacity of 920~mAh~g$^{-1}$ at 100~mA~g$^{-1}$ after 300 cycles \cite{Qiu_HEOLi_JAC_2019}.
(MgCuNiZn)$_{0.65}$Li$_{0.35}$O had an even higher initial discharge capacity of 1930~mAh g$^{-1}$~\cite{Lokcu_HEOBattery_ACSApplMaterInt_2020}.
These values compare very favorably with the current standard graphite anodes, which have a capacity of ${\sim}360$~mAh~g$^{-1}$ \cite{Chen_HEBatteries_ACSApplMaterInt_2021}.

Cation-disordered rocksalt-type oxyfluorides were investigated for use as Li-ion cathodes \cite{Lun_HEBatteries_NatMater_2021},
delivering specific energies of up to 307~mAh g$^{-1}$.
High-entropy oxides based on the layered delafossite $\alpha$-NaFeO$_2$ structures were also examined, but their specific capacities were low compared to state-of-the-art cathode materials \cite{Wang_HEDelafossite_SciRep_2020}.

X-ray absorption spectroscopy was used to investigate the lithiation mechanisms in (MgCoNiCuZn)O \cite{Tavani_HEOlithiation_ChemPhysLett_2020, Ghigna_HEBatteries_ACSApplMaterInt_2020}.
Cu(II) was observed to reduce first to Cu(I) and then to metallic Cu, followed by reduction of Ni and Co to the metallic state.
The high-entropy oxide structure appeared to collapse for charges greater than 400~mAh~g$^{-1}$, and at 800~mAh~g$^{-1}$, a significant portion of Zn metal was present \cite{Ghigna_HEBatteries_ACSApplMaterInt_2020}.
Removing Zn or Mg from (MgCoNiCuZn)O resulted in significantly worse cycling performance, indicating that these cations are important for maintaining the rocksalt structure.

\noindent
\textbf{Catalysts.}
High-entropy catalysts, particularly oxides, demonstrate several advantages over ordered counterparts:
compositional freedom translating to substantial electronic structure and
coordination environment tunability~\cite{Nguyen_AdvSci_2021,Li_NCat_2021},
better performance than fewer-component variants~\cite{Nguyen_AdvSci_2021,Li_NCat_2021},
and excellent stability and durability~\cite{Fracchia_HEOCatalysis_JPhysChemLett_2020,Nguyen_AdvSci_2021,Gao_AppEn_2020,Xu_HECatalysts_NatCommun_2020}.
High entropy enables the design of new catalysts that combine and even enhance beneficial features
of their components.
In FeNiCoCrMn-glycerate, Fe-, Ni-, Co-, and Mn-based materials are known to be highly active oxygen evolution reaction electrocatalysts~\cite{Nguyen_AdvSci_2021},
while Cr induces strain in the structure that weakens active site chemisorption.
The layered glycerate structure allows rapid transport of the reactants to the material and provides additional catalytic active sites.
High entropy can also stabilize exotic active sites:
Cu in high-entropy rocksalt phases can cycle between redox states Cu(II) and Cu(I) as it is treated with CO and O$_{2}$ at temperatures from $130^{\circ}$C to
${\sim250}^{\circ}$C~\cite{Fracchia_HEOCatalysis_JPhysChemLett_2020}, whereas CuO is irreversibly reduced to Cu(I) under the same conditions.

Poly-cation oxides like (FeMgCoNi)O$_{1.2}$ have shown great potential for
hydrogen production through two-step thermochemical water splitting~\cite{Zhai_PCOWaterSplit_EES_2018,Gao_AppEn_2020}.
Improvements have been reported by integrating microwave-absorbing SiC foam
and performing the reaction under short-term, low-energy microwave irradiation:
{\bf i.}~the thermal reduction process was expedited to 4 minutes versus the usual 30 minutes or greater,
{\bf ii.}~high H$_{2}$ generation rates, and
{\bf iii.}~power consumption that is 3\% that of conventional heat treatments.
Microwave irradiation significantly increases the oxygen vacancies in the structure
and can also generate plasma that enhances hydrogen production,
but at the cost of accelerated degradation of the material.

High-entropy structures were also shown to be a catalytic booster for
aerobic oxidative desulfurization of diesel~\cite{Wei_AppSurfSci_2020}.
Carbon and oxygen co-doped hexagonal boron nitride catalysts with the highest density of grain boundaries
demonstrated the shortest induction period, achieving complete sulfur conversion in six hours.

\noindent
\textbf{Thermoelectrics.}
Thermoelectric materials generate a voltage when subjected to a thermal gradient, and are used for energy generation and refrigeration.
They are evaluated by the ``figure of merit'': $zT = \sigma S^2 / \kappa$, where $\sigma$ is the electrical conductivity, $S$ is the Seebeck coefficient and $\kappa$ is the thermal conductivity ---
the reduced lattice thermal conductivity in disordered materials leads to increased $zT$.
The thermal conductivity of perovskite-structure (BaCaLaPbSr)TiO$_3$ was ${\sim}5$ times lower than that of SrTiO$_3$ \cite{Zheng_HEPerovskite_JAdvCeram_2021}, with a maximum $zT$ over 0.2.
The high-entropy chalcogenide Pb$_{0.89}$Sb$_{0.012}$Sn$_{0.1}$Se$_{0.5}$Te$_{0.25}$S$_{0.25}$ had an ultralow thermal conductivity of 0.3~W~m$^{-1}$~K$^{-1}$ \cite{Jiang_HEChalcogenides_Science_2021} with $zT$ of 1.8 compared to 0.8 for Pb$_{0.99}$Sb$_{0.012}$Se (Figure \ref{fig:magnet_battery_thermoelectric_radiation}c).
Adding Sn enabled the stabilization of the Pb(SeTeS) system due to increased configurational entropy.
Pb$_{0.975-x}$Cd$_x$Na$_{0.025}$Se$_{0.5}$S$_{0.25}$Te$_{0.025}$, where $x\leq0.05$, had a conversion efficiency at $\Delta T = 507$~K of 12\% (among the
highest reported values) with a power output of 2.7~W \cite{Jian_HEChalcogenides_NatCommun_2021}.

\noindent
\textbf{Thermal and environmental barriers.}
The simultaneous suppression of the thermal conductivity \cite{Braun_ESO_AdvMat_2018} and enhancement of the mechanical properties  \cite{curtarolo:art140,Ye_Carbide_JAmCerS_2018} in high-entropy materials makes them highly attractive for thermal and environmental protection applications, with recent work focusing on niobates \cite{ Zhu_HENiobates_JEurCeramSoc_2021, Chen_HENiobates_ApplPhysLett_2021} and zirconates \cite{Zhou_HEZirconate_JEurCeramSoc_2020, Zhu_HEZirconate_JEurCeramSoc_2021}.
Pyrochlore-structure (LaNdSmEuGd)$_2$Zr$_2$O$_7$ could withstand up to 5 times as much thermal cycling as La$_2$Zr$_2$O$_7$ \cite{Zhou_HEZirconate_JEurCeramSoc_2020}.
Rare earth niobates had ultra-low thermal conductivities of $<1$~W~m$^{-1}$~K$^{-1}$~\cite{Zhu_HENiobates_JEurCeramSoc_2021} and hardnesses up to $13.9$~GPa \cite{Chen_HENiobates_ApplPhysLett_2021}.

\noindent
\textbf{Nuclear energy applications.}
Their thermochemical stability and mechanical behavior make high-entropy materials promising for nuclear energy applications ranging from {\bf i.} structural components in fission and fusion reactors, which need resistance to neutron irradiation damage, corrosive coolants, thermal and irradiation creep, helium embrittlement, etc. \cite{Wang_IrradiationHEC_ActaMater_2020}; to {\bf ii.} immobilizing matrices for high-level waste, which require excellent radiation tolerance and aqueous durability \cite{Zhou_RadioactiveWastePyrochlore_JHazMater_2021}.
Radiation-induced defect clusters and dislocation loops in (NbTaTiZr)C remained small --- likely due to the lattice distortion slowing their growth --- and no void formation or radiation-induced segregation near grain boundaries were observed  \cite{Wang_IrradiationHEC_ActaMater_2020}.
For pyrochlore-structure  (Eu$_{1-x}$Gd$_x$)$_2$(TiZrHfNbCe)$_2$O$_7$, the normalized leaching rates for simulated radionuclides Ce (surrogate for Pu) and Gd were two orders of magnitude lower than that for Gd$_2$Zr$_2$O$_7$  (Figure \ref{fig:magnet_battery_thermoelectric_radiation}d).
Electrical conductivity was much lower in the high-entropy material, indicating significantly fewer oxygen vacancies --- leading to slower cation diffusion and higher chemical stability --- while lattice distortion also leads to a high lattice potential energy, reducing cation mobility.

\section*{Conclusions}

\noindent
Disorder is propelling the development of new ceramic materials.
Configurational entropy provides new thermodynamic pathways to stabilize multi-component materials.
Disorder directly enhances properties, facilitating the design of materials with features that are traditionally considered mutually exclusive, such as high stiffness and low thermal conductivity,
or improved hardness and fracture toughness.
High-entropy ceramics will continue to impact fields as diverse as batteries, catalysts, thermoelectrics, magnets, ferroelectrics, nuclear energy, and structural materials, for the foreseeable future.

\noindent{\small\textbf{Acknowledgments.}}
The authors thank Xiomara Campilongo, Eva Zurek, William Fahrenholtz, Douglas Wolfe and Darrell Schlom for valuable discussions.
Research sponsored by DOD-ONR (N00014-15-1-2863, N00014-21-1-2515) and NSF (DMR-1921909, DGE-2022040).

\noindent{\small\textbf{Author contributions.}
The authors contributed equally to the article.
}

\noindent{\small\textbf{Competing interests.} The authors declare no competing interests.}

\newcommand{\Ozolins}{Ozoli{\c{n}}{\v{s}}}

\end{document}